%% file: CKM-pen.tex
\documentclass[12pt]{article}
\usepackage{epsfig}

\newcommand{\lsim}{
\mathrel{\hbox{\rlap{\hbox{\lower4pt\hbox{$\sim$}}}\hbox{$<$}}}}

\input econfmacros.tex
\def\Title#1{\begin{center} {\Large {\bf #1} } \end{center}}

\begin{document}

\Title{Penguin Effects in \boldmath$\phi_{d,s}$\unboldmath\ Determinations}

\bigskip\bigskip


\begin{raggedright}  

{\it Robert Fleischer\index{Fleischer, R.}\\
Nikhef, Science Park 105, NL-1098 XG Amsterdam,  NETHERLANDS,\\
Department of Physics and Astronomy, Vrije Universiteit Amsterdam\\
NL-1081 HV Amsterdam, NETHERLANDS}
\bigskip\bigskip
\end{raggedright}

\noindent
{\it {\bf Abstract:} The theoretical precision of the measurement of the 
$B^0_{d,s}$--$\bar B^0_{d,s}$ mixing phases $\phi_{d,s}$ through the benchmark 
decays $B^0_d\to J/\psi K_{\rm S}$, $B^0_s\to J/\psi \phi$ and $B^0_s\to J/\psi f_0(980)$ is 
limited by doubly Cabibbo-suppressed penguin topologies which are usually neglected. 
However, the search for New-Physics
effects in the quark-flavor sector has entered a territory where these effects have to be
taken into account, which will be particularly relevant for the LHCb upgrade era. 
Thanks to their non-perturbative nature, the penguin corrections cannot be calculated 
but have rather to be controlled through experimental data. An overview of the picture of 
the penguin parameters originating from the current data for $B_{(s)}\to J/\psi \pi, J/\psi K$
decays and the physics potential of the control channels $B^0_s\to J/\psi K_{\rm S}$,
$B^0_s\to J/\psi \bar K^{*0}$ and $B_d^0\to J/\psi f_0(980)$ is given, emphasizing also
the usefulness of effective $B_s$ decay lifetimes. 

\vspace*{0.7truecm}

\noindent 
Proceedings of CKM 2012, the 7th International Workshop on the CKM Unitarity Triangle, University of Cincinnati, USA, 28 September -- 2 October 2012}

\section{Introduction}
This summer, we have received the great news that a Higgs-like particle was 
discovered by the ATLAS and CMS collaborations at the LHC. On the other hand, 
these experiments could still not reveal any deviation from the Standard Model (SM) 
at the high-energy frontier, while the LHCb experiment operating at the high-precision 
frontier could also not yet resolve evidence for New Physics (NP) in the quark-flavor 
sector. Concerning the structure of possible physics lying beyond the SM, we have therefore
to deal with a larger characteristic NP scale $\Lambda_{\rm NP}$, i.e.\ not just 
$\Lambda_{\rm NP}\sim \mbox{TeV}$,  or/and symmetries prevent large NP effects 
in the flavor sector, where the most prominent example is given by models 
with ``Minimal Flavor Violation".

Many more results are yet to come, but in view of the current situation we have to
prepare ourselves to deal with smallish NP effects. In order to resolve such phenomena,
it is crucial to have a critical look at theoretical analyses and the approximations involved. The 
central issue is related to strong interactions and ``hadronic" uncertainties. In particular, the 
theoretical and experimental precisions have to be matched to one another, which will be 
especially relevant for the LHCb upgrade program.

Concerning the further exploration of CP violation, $B_d\to J/\psi K_{\rm S}$, 
$B_s\to J/\psi \phi$ and $B_s\to J/\psi f_0(980)$ decays play outstanding roles, allowing
measurements of the $B^0_{d,s}$--$\bar B^0_{d,s}$ mixing phases 
\begin{equation}
\phi_d=2\beta + \phi_d^{\rm NP}, \quad \phi_s=-2\delta\gamma + \phi_s^{\rm NP}.
\end{equation}
Here the former pieces are the SM contributions 
$\phi_q^{\rm SM}=2\mbox{arg}(V_{tq}^*V_{tb})$, with $\beta$ denoting the usual angle of the
CKM unitarity triangle, and $\delta\gamma\approx1^\circ$ \cite{RF-rep}. From 
the theoretical point of view, these measurements are affected by uncertainties from doubly 
Cabibbo-suppressed penguin contributions \cite{RF-BsJpsiK}--\cite{MJ}. These effects 
are usually neglected in the experimental analyses and are naively expected to be very small.
However, as they are related to non-perturbative long-distance dynamics, the corresponding
parameters cannot be calculated within perturbative QCD. Consequently, the question arises
how big these effects are and how they can be controlled by means of experimental data.

\section{\boldmath$B^0_d\to J/\psi K_{\rm S}$  and 
$B^0_{s}\to J/\psi K_{\rm S}$\unboldmath}\label{sec:BsJpsiK}
In the SM, the $B_d^0\to J/\psi\, K_{\rm S}$ decay amplitude can be written as follows 
\cite{RF-BsJpsiK}:
\begin{equation}\label{BJpsiK-ampl}
A(B_d^0\to J/\psi\, K_{\rm S})=\left(1-\lambda^2/2\right){\cal A'}
\left(1+\epsilon a'e^{i\theta'}e^{i\gamma}\right).
\end{equation}
Here $\lambda\equiv |V_{us}|$ is the Wolfenstein parameter, $\gamma$ denotes the usual
angle of the CKM unitarity triangle, and the following CP-conserving 
hadronic parameters enter:
\begin{equation}\label{hadr}
{\cal A}'\equiv \lambda^2 A \left[A_{{\rm T}}^{(c)'}+A_{{\rm P}}^{(c)'}-
A_{{\rm P}}^{(t)'}\right], \quad a' e^{i\theta'}\equiv R_b
\left[\frac{A_{{\rm P}}^{(u)'}-A_{{\rm P}}^{(t)'}}{A_{{\rm T}}^{(c)'}+
A_{{\rm P}}^{(c)'}-A_{{\rm P}}^{(t)'}}\right],
\end{equation}
where $A_{{\rm T}}^{(c)'}$ is the color-suppressed tree contribution
and the $A_{{\rm P}}^{(q)'}$  denote penguin topologies with internal $q$ quarks. The primes
remind us that we are dealing with a $\bar b\to \bar cc \bar s$  transition. Moreover,
the decay amplitude involves the CKM factors 
\begin{equation}
A\equiv \frac{1}{\lambda^2}|V_{cb}|\sim 0.8, \quad 
R_b\equiv \left(1-\frac{\lambda^2}{2}\right)\frac{1}{\lambda}
\left|\frac{V_{ub}}{V_{cb}}\right|\sim0.5, \quad 
\epsilon\equiv\frac{\lambda^2}{1-\lambda^2}=0.053.
\end{equation}
The  parameters in (\ref{hadr}) suffer from large hadronic uncertainties, in particular
the $a'e^{i\theta'}$, which measures the ratio of tree to penguin contributions. However, 
as the latter quantity is doubly Cabibbo-suppressed in (\ref{BJpsiK-ampl}) by 
the tiny $\epsilon$,  it is usually neglected. 

The $B_d\to J/\psi\, K_{\rm S}$ channel offers the following time-dependent CP asymmetry:
\begin{eqnarray}
\lefteqn{\frac{\Gamma(B^0_d(t)\to J/\psi K_{\rm S})-
\Gamma(\bar B^0_d(t)\to J/\psi K_{\rm S})}{\Gamma(B^0_d(t)\to J/\psi K_{\rm S})+
\Gamma(\bar B^0_d(t)\to J/\psi K_{\rm S})}}\nonumber\\
&&=C(B_d\to J/\psi K_{\rm S})\cos(\Delta M_d t)-S(B_d\to J/\psi K_{\rm S})\sin(\Delta M_d t),
\end{eqnarray}
where 
\begin{equation}
C(B_d\to J/\psi K_{\rm S})=
-\frac{2\epsilon a'\sin\theta'\sin\gamma}{1+2\epsilon a'\cos\theta'\cos\gamma+\epsilon^2a'^2}
\end{equation}
describes direct CP violation, and the ``mixing-induced" CP asymmetry
\begin{equation}\label{S-def}
\frac{S(B_d\to J/\psi K_{\rm S})}{\sqrt{1-C(B_d\to J/\psi K_{\rm S})^2}}
=\sin(\phi_d+\Delta\phi_d)
\end{equation}
originates from the interference between $B^0_d$--$\bar B^0_d$ mixing and decay processes. 
The phase shift $\Delta\phi_d$ is given by the following expression \cite{FFJM}:
\begin{equation}\label{tanDel}
\tan \Delta\phi_d =\frac{2 \epsilon a'\cos\theta' \sin\gamma+\epsilon^2a'^2
\sin2\gamma}{1+ 2 \epsilon a'\cos\theta' \cos\gamma+\epsilon^2a'^2\cos2\gamma}.
\end{equation}

The most recent Heavy Flavor Averaging Group (HFAG) compilation \cite{HFAG} gives
\begin{equation}
C(B_d\to J/\psi K_{\rm S})= 0.024 \pm 0.026 \, \Rightarrow \, 
\sqrt{1-C(B_d\to J/\psi K_{\rm S})^2}=0.9997^{+0.0003}_{-0.0010},
\end{equation}
so that (\ref{S-def}) can be simplified with excellent precision as
\begin{equation}\label{S-exp}
S(B_d\to J/\psi K_{\rm S})=\sin(\phi_d+\Delta\phi_d)=0.665 \pm 0.024.
\end{equation}
This expression illustrates the theoretical limitation of the measurement of $\phi_d$
through the phase shift $\Delta\phi_d$, which is caused by the doubly Cabibbo-suppressed
penguin contributions, as can be seen in (\ref{tanDel}). For values of $a'\sim0.2$, the 
$\Delta\phi_d$ can be as large as about $1^\circ$, depending on the strong phase 
$\theta'$.

How can we control $\Delta\phi_d$? As $a'$ and $\theta'$
cannot be calculated reliably, we use the control channel $B^0_s\to J/\psi K_{\rm S}$, 
which is caused by $\bar b\to \bar c c\bar d$ quark-level processes and is
related to $B^0_d\to J/\psi K_{\rm S}$ through the $U$-spin flavor symmetry of strong
interactions \cite{RF-BsJpsiK}. Its decay amplitude can be written in the SM as
\begin{equation}\label{BsJpsiK}
A(B_s^0\to J/\psi\, K_{\rm S})=-\lambda\,{\cal A}\left(1-a e^{i\theta}e^{i\gamma}\right),
\end{equation}
where the unprimed amplitudes are defined in analogy to their counterparts in (\ref{hadr}). 
The $U$-spin symmetry implies $a=a'$ and $\theta=\theta'$. The key feature of (\ref{BsJpsiK})
is the absence of the $\epsilon$ suppression factor in front of the $a e^{i\theta}$. 
Consequently, the impact of this parameter is magnified in the corresponding observables. 

As was pointed out in \cite{RF-BsJpsiK}, $\gamma$ as well as $a$ and $\theta$ 
can be determined from the CP asymmetries of $B_s\to J/\psi K_{\rm S}$ 
and the ratio of the $B_d\to J/\psi K_{\rm S}$, $B_s\to J/\psi K_{\rm S}$  branching ratios.
While the $\gamma$ determination appeared most  interesting back in 1999, there has been 
a recent change of focus  \cite{DeBFK}: the extraction of $\gamma$ looks feasible 
at the LHCb upgrade but probably not competitive with other methods. However, using 
$\gamma$ as an input, the hadronic parameters $a$, $\theta$ can be determined in a 
{\it clean} way from the CP-violating $B_s\to J/\psi K_{\rm S}$ observables, thereby allowing 
us to get a handle on the penguin effects in the measurement of $\phi_d$ from 
$S(B_d\to J/\psi K_{\rm S})$. 

The $B_s\to J/\psi K_{\rm S}$ channel has been observed by CDF \cite{CDF-BsJpsiKS}
and LHCb \cite{LHCb-BsJpsiKS}, but so far only measurements of the branching ratio 
are available, where subtleties related to the sizable $B_s$ decay width difference 
$\Delta\Gamma_s$ have to be taken into account \cite{BR-paper}. These branching
ratio measurements are consistent with an $SU(3)$ relation to the branching ratio of
$B_d\to J/\psi \pi^0$ \cite{DeBFK,LHCb-BsJpsiKS}. It is useful to introduce the ratio
\begin{equation}\label{H-def}
H\equiv \frac{1}{\epsilon}
\left|\frac{{\cal A}'}{{\cal A}}\right|^2
\left[\frac{\tau_{B_d}\Phi^d_{J/\psi K_{\rm S}}}{\tau_{B_s}\Phi^s_{J/\psi K_{\rm S}}}\right]
\frac{{\rm BR}
\left(B_s \to J/\psi K_{\rm S}\right)}{{\rm BR}(B_d\to J/\psi K_{\rm S})},
\end{equation}
where the $\Phi$ and $\tau_{B_q}$ denote phase-space factors and $B_q$ lifetimes, respectively. 

\begin{figure}[tbp] 
   \centering
   \includegraphics[width=2.5in]{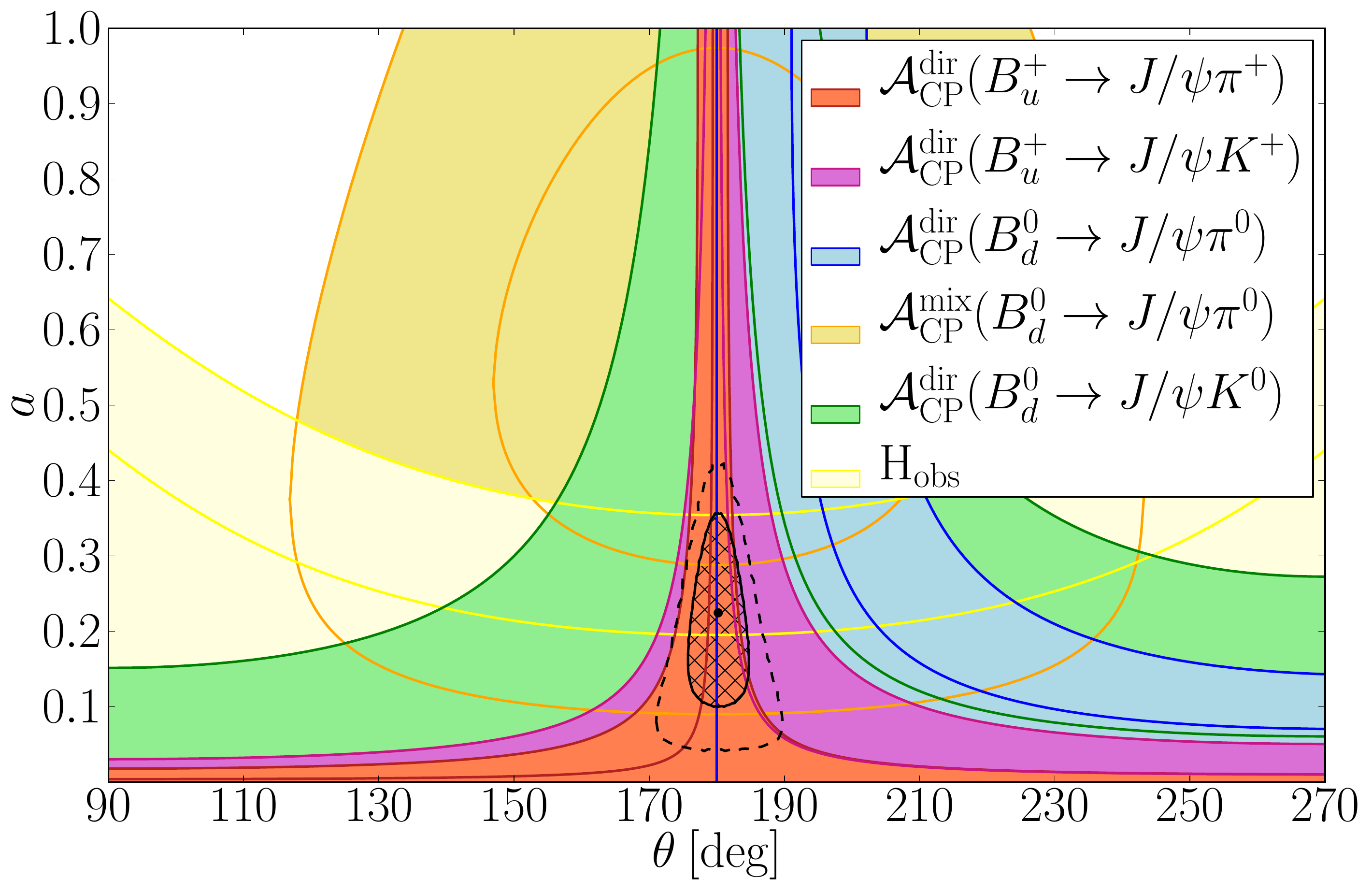} ~~
    \includegraphics[width=2.6in]{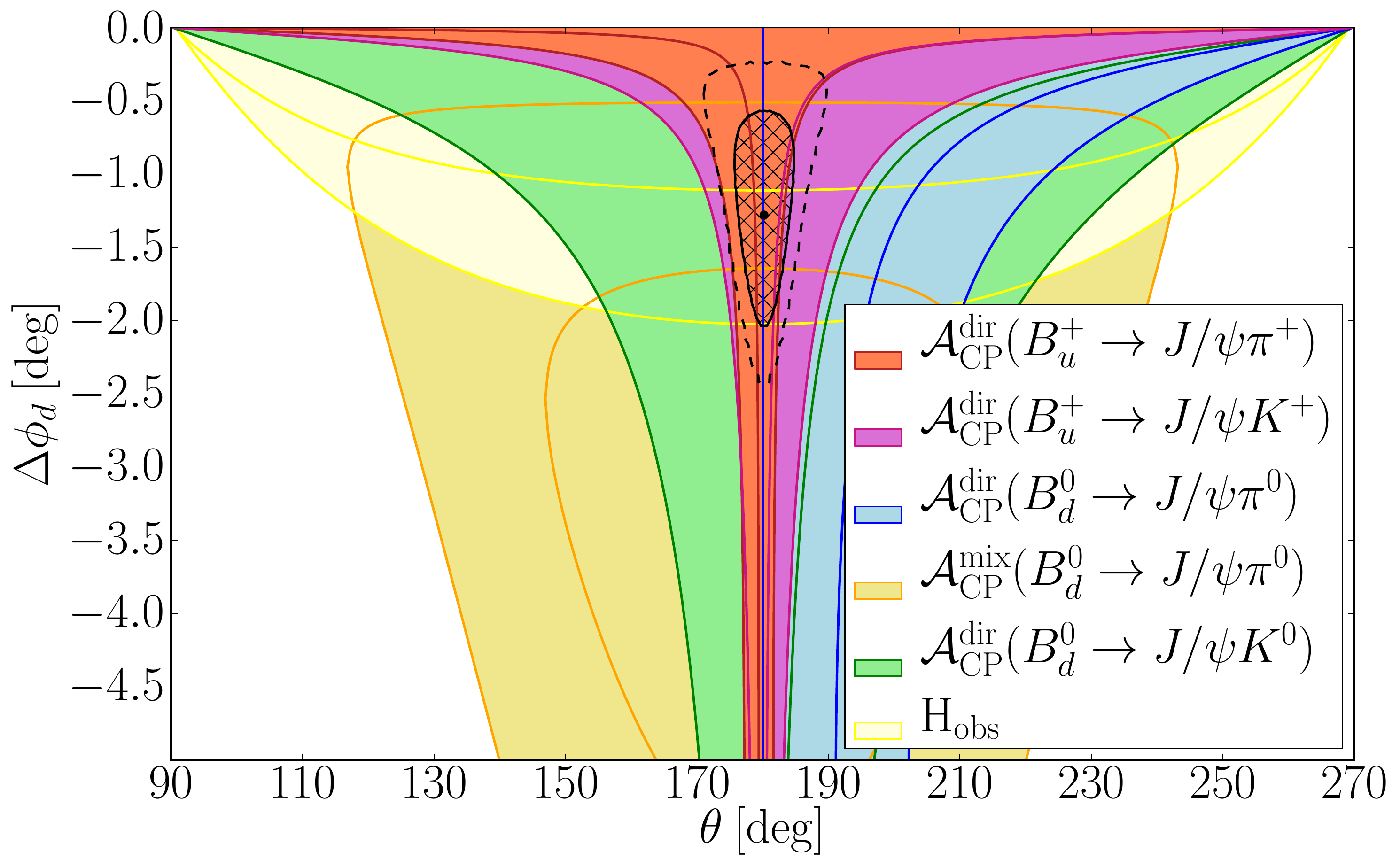} 
   \caption{Current experimental constraints for the penguin parameters $a$, $\theta$ (left) and the
    phase shift $\Delta\phi_d$ (right), showing the 39\% and 68\% C.L. regions (from \cite{DeBF}).}
   \label{fig:pen-const}
\end{figure}

In order to constrain  $a$, $\theta$ from currently available 
data, we use also decays with a CKM structure similar to $B^0_s\to J/\psi K_{\rm S}$, i.e.\
$B^0_d\to J/\psi \pi^0$ and $B^+\to J/\psi \pi^+$, and complement them with 
$B^0_d\to J/\psi K^0$, $B^+\to J/\psi K^+$ decays. These channels allow 
the construction of a variety of $H$ ratios in analogy to (\ref{H-def}). The data 
give an internally consistent picture, with the average 
$H_{\rm obs}=1.19\pm0.04 {\rm (stat)}\pm 0.21 {\rm (FF)}$, which 
takes also $SU(3)$-breaking corrections through form-factor (FF) ratios into account 
\cite{DeBF}. In Fig.~\ref{fig:pen-const}, the current picture 
is shown, corresponding to the following $1\sigma$ ranges:
\begin{equation}\label{pen-par}
a=0.22 \pm 0.13, \quad \theta=(180.2\pm4.5)^\circ, \quad \Delta\phi_d=-(1.28\pm0.74)^\circ.
\end{equation}

The situation of the analysis and extraction of the penguin parameters for the LHCb upgrade
looks promising \cite{DeBFK,DeBF}, where the $B_s\to J/\psi K_{\rm S}$ channel is expected
to play the role of a golden mode to explore the importance of penguin topologies.

\section{\boldmath$B^0_s\to J/\psi \phi$ and $B^0_s\to J/\psi \bar K^{*0}$\unboldmath}
The decay $B^0_s\to J/\psi \phi$ is the $B_s$-meson counterpart of the 
$B^0_d\to J/\psi K_{\rm S}$ channel, allowing the extraction of the $B^0_s$--$\bar B^0_s$
mixing phase $\phi_s$. Since the final state involves two vector mesons, a time-dependent
angular analysis has to be performed in order to disentangle the CP eigenstates \cite{DDF}.
In analogy to $B^0_d\to J/\psi K_{\rm S}$, the analysis of CP violation in the 
$B^0_s\to J/\psi \phi$ channel is also affected by doubly Cabibbo-suppressed penguin
contributions \cite{FFM}. For a given final-state configuration $f\in\{0,\parallel,\perp\}$,
the SM decay amplitude can be written as
\begin{equation}
A(B^0_s\to (J/\psi \phi)_f)=\left(1-\lambda^2/2\right)
 {\cal A}_f'\left[1+ \epsilon\, a_f' e^{i\theta_f'}{e^{i\gamma}}\right], 
\end{equation}
and the mixing-induced CP asymmetries take the form
\begin{equation}\label{ACP-mix}
{\cal A}_{{\rm CP}}^{\rm mix}(B_s\to (J/\psi \phi)_f)=\sin(\phi_s+\Delta\phi_s^f),
\end{equation}
which is the counterpart of (\ref{S-exp}). In the literature, it is usually assumed that
$\Delta\phi_s^f=0$. The most recent average compiled by HFAG reads 
$\phi_s=-(0.74^{+5.2}_{-4.8})^\circ$ \cite{HFAG}, whereas we have 
$\phi_s=-(2.08\pm0.09)^\circ$ in the SM \cite{CKM-fitter}. In view of the small value of
$\phi_s$ emerging from the data, a phase shift $\Delta \phi_s^f$ at the $1^\circ$ level (see
(\ref{pen-par})) would have a significant impact for the resolution of possible NP effects. 

A channel to probe these penguin contributions is offered by $B^0_s\to J/\psi \bar K^{*0}$, with
a SM decay amplitude of the structure
\begin{equation}
A(B_s^0\to (J/\psi \bar K^{*0})_f)=-\lambda\,{\cal A}_f\left[1-a_f e^{i\theta_f} e^{i\gamma}\right],
\end{equation}
which is similar to (\ref{BsJpsiK}). In particular, $a_f e^{i\theta_f}$ is again not suppressed by the 
tiny $\epsilon$ parameter. Neglecting penguin annihilation ($PA$) and exchange 
topologies $(E)$, which can be constrained by the upper bound on BR$(B_d\to J/\psi\phi)$ as
 $|E+PA|/|T|\lsim0.1$, and using the $SU(3)$ flavor symmetry, 
we get the relations $a_f=a_f'$ and $\theta_f=\theta_f'$, allowing us to determine/constrain 
the penguin shift $\Delta\phi_s^f$ in (\ref{ACP-mix}) \cite{FFM}. 

In contrast to $B^0_s\to J/\psi K_{\rm S}$, $B^0_s\to J/\psi \bar K^{*0}$ is 
flavor-specific and does, hence, not show mixing-induced CP violation. Consequently, the 
implementation of this method has to use measurements of 
untagged and direct CP-violating observables, and an angular analysis is required 
to disentangle the final-state configurations $f$.

The $B^0_s\to J/\psi \bar K^{*0}$ decay was observed by CDF \cite{CDF-BsJpsiKS}
and LHCb \cite{LHCb-BsKast}. The most recent LHCb branching ratio 
$(4.4^{+0.5}_{-0.4}\pm0.8) \times 10^{-5}$ agrees well with the prediction 
$(4.6\pm0.4) \times 10^{-5}$ obtained from the BR$(B_d\to J/\psi \rho^0)$ by means
of the $SU(3)$ flavor symmetry \cite{FFM}, and the polarization fractions agree well with 
those of $B^0_d\to J/\psi K^{*0}$. 

The experimental sensitivity for the extraction of $\phi_s$ from $B_s\to J/\psi \phi$ 
at the LHCb upgrade (50\,$\mbox{fb}^{-1}$) is expected as 
$\Delta\phi_s|_{\rm exp}\sim 0.008 = 0.46 ^\circ$ \cite{LHCb-Strat}. In view of this impressive
precision on the one hand and $\Delta\phi_d=-(1.28\pm0.74)^\circ$ following from the current 
data for $B \to J/\psi \pi, J/\psi K$ decays with a dynamics similar to $B_s\to J/\psi \phi$ (see 
Section~\ref{sec:BsJpsiK}) on the other hand, it will be crucial to get a handle on the penguin 
effects at the LHCb upgrade as they may mimic NP effects. 

\section{\boldmath$B^0_s\to J/\psi f_0(980)$ and $B^0_d\to J/\psi f_0(980)$\unboldmath}
Another decay that has recently entered the stage is $B^0_s\to J/\psi f_0(980)$,
which was observed by LHCb \cite{LHCb-f0}, Belle \cite{Belle-f0}, CDF \cite{CDF-f0} and
D0 \cite{D0-f0}. The dominant
decay mode is via $f_0\to\pi^+\pi^-$, with a branching ratio about four times smaller than
that of $B_s^0\to J/\psi \phi$ with $\phi\to K^+ K^-$. However, since the $f_0\equiv f_0(980)$ 
is a scalar $J^{PC}=0^{++}$ state no angular analysis is required, thereby simplifying the analysis
considerably and offering an interesting alternative for the determination of $\phi_s$ \cite{SZ}. 

The impact of hadronic uncertainties on the extraction of $\phi_s$ from CP violation in
$B^0_s\to J/\psi f_0$ was studied in detail in \cite{FKR}, and for the 
$B_{s,d}\to J/\psi \eta^{(\prime)}$ system in \cite{FKR-eta}. The general formalism is very 
similar to the discussion given above:
\begin{equation}
A(B^0_s\to J/\psi f_0)\propto
\left [1+\epsilon b e^{i\vartheta} e^{i\gamma}  \right],
\end{equation}
i.e.\ the hadronic penguin corrections enter again in a doubly Cabibbo-suppressed way. 
The mixing-induced CP asymmetry can be written as
\begin{equation}
S(B^0_s\to J/\psi f_0)=\sqrt{1-C(B^0_s\to J/\psi f_0)^2}\sin(\phi_s+\Delta\tilde\phi_s),
\end{equation}
where $\Delta\tilde\phi_s$ is given by an expression analogous to (\ref{tanDel}). 
However, in contrast to the $B_d\to J/\psi K_{\rm S}$ and $B_s\to J/\psi \phi$ decays, 
the $B_s\to J/\psi f_0$ channel suffers from the fact that the hadronic structure of the 
$f_0(980)$ is poorly known: popular benchmark scenarios are the quark--antiquark 
and tetraquark pictures. In the latter case, a peculiar decay topology arises at the tree level 
that does not have a counterpart in the quark--antiquark description \cite{FKR}. 

The parameter $b$ depends on the hadronic composition of the $f_0$ and is therefore
essentially unknown. Making the conservative assumption $0\leq b\leq0.5$ (where the upper 
bound of $0.5$ is related to the $R_b\sim0.5$ factor in (\ref{hadr})) and 
$0^\circ\leq\vartheta\leq360^\circ$ yields $\Delta\tilde\phi_s\in [-2.9^\circ, 2.8^\circ]$.
This range translates into the SM range
\begin{equation}\label{f0-SM}
\left.S(B_s \to J/\psi f_0)\right|_{\rm SM} \in [ -0.086, -0.012],
\end{equation}
while the naive value with $\Delta\tilde\phi_s=0^\circ$ 
reads  $(\sin\phi_s)|_{\rm SM}=-0.036\pm 0.002$ \cite{FKR}.

An alternative to determine the $B^0_s$--$\bar B^0_s$ mixing parameters is 
offered by effective $B_s$ decay lifetimes \cite{FK,RK}, which are defined for 
a general $B_s\to f$ decay as
\begin{equation}
  \tau_{f} 
  \equiv \frac{\int^\infty_0 t\  [\Gamma(B^0_s(t)\to f) + \Gamma(\bar B^0_s(t)\to f)]\ dt}
  {\int^\infty_0 [\Gamma(B^0_s(t)\to f) + \Gamma(\bar B^0_s(t)\to f)]\ dt}.
\end{equation}
The effective lifetimes of $B_s$ decays into CP-even (such as $B_s\to K^+K^-$) 
and CP-odd (such as $B_s \to J/\psi f_0$) final states allow the extraction of
$\phi_s$ and the decay width difference $\Delta\Gamma_s$. This 
determination is extremely robust with respect to the hadronic penguin uncertainties, thereby
nicely complementing studies of CP violation. First experimental results are already available
\cite{Dordei}, and future lifetime measurements with $1\%$ uncertainty would be most
interesting. 

The current LHCb result for the extraction of $\phi_s$ from $B_s \to J/\psi f_0$ is given by
$\phi_s=-(25\pm25\pm1)^\circ$, which corresponds to $S=-0.43^{+0.43}_{-0.34}$ \cite{Raven}.
In this analysis, hadronic corrections were not taken into account and are not yet relevant
in view of the large experimental errors. However, once the data will enter the SM range 
in (\ref{f0-SM}), we have to start to constrain the $\Delta\tilde\phi_s$. Since the hadronic 
effects have a different impact on $B^0_s\to J/\psi f_0$ and $B^0_s\to J/\psi \phi$, it will 
be interesting to compare the individual measurements of CP violation. 

A way to obtain insights into the   penguin effects is offered 
by $B^0_d\to J/\psi f_0$. Its branching ratio with $f_0\to\pi^+\pi^-$
could be as large as ${\cal O}(10^{-6})$ \cite{FKR}.
The translation of the corresponding penguin parameters into those
of $B_s \to J/\psi f_0$ depends unfortunately also on assumptions about the hadronic structure
of the $f_0(980)$. By the time these measurements may become available we will hopefully
also have a better picture of this scalar hadronic state.

\section{Conclusions}\label{sec:concl}
We are currently moving towards new frontiers in terms of precision. Despite the observation
of a Higgs-like new particle, the LHC has not yet revealed signals of physics beyond the 
SM. Consequently, we have to prepare ourselves to deal with smallish NP effects, 
matching in particular the steadily  increasing experimental precision of $B$-decay
studies with the precision of the corresponding theoretical analyses.

In the case of the determination of the $\phi_{d,s}$ mixing phases from the benchmark
decays, we are entering a territory where doubly Cabibbo-suppressed penguin contributions,
which could so far be neglected, have to be controlled. 
The currently available data for $B_{(s)}\to J/\psi \pi, J/\psi K$ decays give $a=0.22 \pm 0.13$, 
$\theta=(180.2\pm4.5)^\circ$  with a phase shift of 
$\Delta\phi_d=-(1.28\pm0.74)^\circ$, thereby setting the scale of the penguin effects and
the associated uncertainties. 

At the LHCb upgrade, the $B^0_s\to J/\psi K_{\rm S}$ decay will play
an outstanding role for exploring these effects. Further insights for the measurement of
$\phi_s$ from $B^0_s\to J/\psi \phi$ can be obtained from $B^0_s\to J/\psi \bar K^{*0}$. In the
case of the $B_{s,d}\to J/\psi f_0(980)$ system, the hadronic structure of the $f_0(980)$ affects 
the uncertainty of the corresponding value of $\phi_s$. Future measurements of effective $B_s$
decay lifetimes with precisions at the $1\%$ level would offer interesting alternatives for the 
extraction of $\phi_s$, which are very robust with respect to hadronic uncertainties.

\bigskip
{\it Acknowledgement:} I am grateful to Kristof De Bruyn for discussions, updates and 
numerical work related to Section~\ref{sec:BsJpsiK}, and would also like to thank 
my other colleagues involved in the topics discussed above for the enjoyable collaboration.

\end{document}

%% file: econfmacros.tex
\def\beq{\begin{equation}}
\def\eeq#1{\label{#1}\end{equation}}
\def\eeqn{\end{equation}}

\def\beqa{\begin{eqnarray}}
\def\eeqa#1{\label{#1}\end{eqnarray}}
\def\eeqan{\end{eqnarray}}

\let\bar=\overbar

\def\Dslash{\not{\hbox{\kern-4pt $D$}}}
\def\dslash{\not{\hbox{\kern-2pt $\del$}}}

\def\msb{{\bar{\ssstyle M \kern -1pt S}}}




%% file: CKM-pen.bbl
\begin{thebibliography}{99}

\bibitem{RF-rep}R.~Fleischer,
  Phys.\ Rept.\  {\bf 370}, 537 (2002)
  [hep-ph/0207108].

\bibitem{RF-BsJpsiK}R.~Fleischer,
  Eur.\ Phys.\ J.\ C {\bf 10}, 299 (1999)
  [hep-ph/9903455].

\bibitem{CPS}M.~Ciuchini, M.~Pierini and L.~Silvestrini,
  Phys.\ Rev.\ Lett.\  {\bf 95}, 221804 (2005) [hep-ph/0507290];
  arXiv:1102.0392 [hep-ph].

\bibitem{FFJM}S.~Faller, R.~Fleischer, M.~Jung  and T.~Mannel,
  Phys.\ Rev.\ D {\bf 79}, 014030 (2009)
  [arXiv:0809.0842 [hep-ph]].
  
\bibitem{FFM}S.~Faller, R.~Fleischer and T.~Mannel,
  Phys.\ Rev.\ D {\bf 79}, 014005 (2009)
  [arXiv:0810.4248 [hep-ph]].
    
\bibitem{DeBFK}K.~De Bruyn, R.~Fleischer and P.~Koppenburg,
  Eur.\ Phys.\ J.\ C {\bf 70}, 1025 (2010)
  [arXiv:1010.0089 [hep-ph]].
  
\bibitem{FKR}R.~Fleischer, R.~Knegjens and G.~Ricciardi,
  Eur.\ Phys.\ J.\ C {\bf 71}, 1832 (2011)
  [arXiv:1109.1112 [hep-ph]].
  
\bibitem{MJ}M.~Jung,
  Phys.\ Rev.\ D {\bf 86}, 053008 (2012)
  [arXiv:1206.2050 [hep-ph]].
  
\bibitem{HFAG}Y.~Amhis {\it et al.}  [Heavy Flavor Averaging Group Collaboration],
  arXiv:1207.1158 [hep-ex]; for online updates, see http://www.slac.stanford.edu/xorg/hfag/

\bibitem{CDF-BsJpsiKS}T.~Aaltonen {\it et al.}  [CDF Collaboration],
  Phys.\ Rev.\ D {\bf 83}, 052012 (2011)
  [arXiv:1102.1961 [hep-ex]].

\bibitem{LHCb-BsJpsiKS}R. Aaij {\it et al.}  [LHCb Collaboration],
  Phys.\ Lett.\ B {\bf 713}, 172 (2012)
  [arXiv:1205.0934 [hep-ex]].

\bibitem{BR-paper}K.~De Bruyn, R.~Fleischer, R.~Knegjens, P.~Koppenburg, M.~Merk 
and N.~Tuning,
  Phys.\ Rev.\ D {\bf 86}, 014027 (2012) 
  [arXiv:1204.1735 [hep-ph]].

\bibitem{DeBF}K.~De Bruyn, R.~Fleischer and P. Koppenburg, in preparation.

\bibitem{DDF}A.~S.~Dighe, I.~Dunietz and R.~Fleischer,
  Eur.\ Phys.\ J.\ C {\bf 6}, 647 (1999)
  [hep-ph/9804253].
  
\bibitem{CKM-fitter}J.~Charles, O.~Deschamps, S.~Descotes-Genon, R.~Itoh, H.~Lacker, A.~Menzel, S.~Monteil and V.~Niess {\it et al.},
  Phys.\ Rev.\ D {\bf 84}, 033005 (2011)
  [arXiv:1106.4041 [hep-ph]].
  
\bibitem{LHCb-BsKast}R. Aaij {\it et al.}  [LHCb Collaboration],
  Phys.\ Rev.\ D {\bf 86}, 071102 (2012)
  [arXiv:1208.0738 [hep-ex]].

\bibitem{LHCb-Strat}R. Aaij {\it et al.}  [LHCb Collaboration] and A. Bharucha {\it et al.},
  arXiv:1208.3355 [hep-ex].
  
\bibitem{LHCb-f0}R.~Aaij {\it et al.}  [LHCb Collaboration],
  Phys.\ Lett.\  B {\bf 698}, 115 (2011) 
  [arXiv:1102.0206 [hep-ex]].
 
\bibitem{Belle-f0}J.~Li {\it et al.}  [Belle Collaboration],
  Phys.\ Rev.\ Lett.\  {\bf 106}, 121802 (2011) 
  [arXiv:1102.2759 [hep-ex]].
 
\bibitem{CDF-f0}T.~Aaltonen {\it et al.}  [CDF Collaboration],
  Phys.\ Rev.\ D {\bf 84}, 052012 (2011)
  [arXiv:1106.3682 [hep-ex]].
  
\bibitem{D0-f0}V.~M.~Abazov {\it et al.}  [D0 Collaboration],
  Phys.\ Rev.\ D {\bf 85}, 011103 (2012)
  [arXiv:1110.4272 [hep-ex]].

\bibitem{SZ}S.~Stone and L.~Zhang,
  Phys.\ Rev.\  D {\bf 79}, 074024 (2009)  [arXiv:0812.2832 [hep-ph]];
  arXiv:0909.5442 [hep-ex].

\bibitem{FKR-eta}R.~Fleischer, R.~Knegjens and G.~Ricciardi,
  Eur.\ Phys.\ J.\ C {\bf 71}, 1798 (2011)
  [arXiv:1110.5490 [hep-ph]].
  
\bibitem{FK}R.~Fleischer and R.~Knegjens,
  Eur.\ Phys.\ J.\ C {\bf 71}, 1789 (2011)
  [arXiv:1109.5115 [hep-ph]].

\bibitem{RK}R.~Knegjens,
  arXiv:1209.3206 [hep-ph].

\bibitem{Dordei}F. Dordei, these proceedings. 

\bibitem{Raven}G. Raven, these proceedings. 
\end{thebibliography}
